# Direct Visualization of the Nematic Superconductivity in $Cu_xBi_2Se_3$

Ran Tao[1], Ya-Jun Yan[1*], Xi Liu[1], Zhi-Wei Wang[3], Yoichi Ando[3], Qiang-Hua Wang[4,5], Tong Zhang[1,2], Dong-Lai Feng[1,2*]

[1] State Key Laboratory of Surface Physics, Department of Physics, and Advanced Materials Laboratory, Fudan University, Shanghai 200438, China
[2] Collaborative Innovation Center of Advanced Microstructures, Fudan University, Shanghai 200438, China
[3] Physics Institute II, University of Cologne, 50937 Cologne, Germany
[4] National Laboratory of Solid State Microstructures & School of Physics, Nanjing University, Nanjing, 210093, China
[5] Collaborative Innovation Center of Advanced Microstructures, Nanjing 210093, China

*Email: yanyajun@fudan.edu.cn, dlfeng@fudan.edu.cn

**$Cu_xBi_2Se_3$ hosts both topological surface states and bulk superconductivity. It has been identified recently as a topological superconductor (TSC) with an extraordinary nematic, i. e. $C_2$-symmetric, superconducting state and odd-parity pairing. Here, using scanning tunneling microscopy (STM), we directly examine the response of the superconductivity of $Cu_xBi_2Se_3$ to magnetic field. Under out-of-plane fields ($B_\perp$), we discover elongated magnetic vortices hosting zero-bias conductance peaks consistent with the Majorana bound states expected in a TSC. Under in-plane fields ($B_{//}$), the average superconducting gap exhibits two-fold symmetry with field orientation; the long $C_2$ symmetry axes are pinned to the dihedral mirror planes under $B_{//}$=0.5 T but rotate slightly under $B_{//}$=1.0 T. Moreover, a nodeless $\Delta_{4x}$ gap structure is semi-quantitatively determined for the first time. Our data paint a microscopic picture of the nematic superconductivity in $Cu_xBi_2Se_3$ and pose strong constraints on theory.**

## I. INTRODUCTION

Topological superconductors, as host of Majorana fermions and Majorana zero modes (MZM) [1], may facilitate topological quantum computing. However, in practice, TSCs are rare. Among various recipes for making a TSC [1,2], bulk superconductors that host topological surface states are the most natural candidates. However, unlike the theoretical triumph in predicting topological insulators [1,3], metals, and semi-metals, theory and experiment diverge when it comes to TSCs. Consequently, there is a large gap in our understanding of TSCs, and in particular, details of the microscopic behavior of TSCs are in urgent demand for improving theory.

$Cu_xBi_2Se_3$ is a prototypical example of the difficulties encountered. Since it hosts both topologically non-trivial surface states and bulk superconductivity [4,5], it has been proposed to be a TSC, likely with an odd-parity pairing symmetry [6]. However, experimental results on $Cu_xBi_2Se_3$ present challenges for this interpretation. The absence of Pauli limiting behavior in the upper critical field suggests spin-triplet superconductivity or an anisotropic spin-singlet state [7], and point-contact spectroscopy found a zero-bias conductance peak (ZBCP) on a cleaved surface [8], which was attributed to MZM. However, Andreev reflection spectroscopy on $Cu_xBi_2Se_3$ shows that the existence of the ZBCP depends on the barrier strength – its absence under finite barrier strength implies the absence of zero-energy Majorana fermions [9]. Furthermore, a low-temperature STM study observed a full superconducting gap without in-gap states, nor any ZBCP in the vortex core, discrediting $Cu_xBi_2Se_3$ as a TSC [10].

Recently, a nematic superconducting state was discovered in an NMR study of $Cu_xBi_2Se_3$ (Ref. [11]). The spin susceptibility in the superconducting state exhibits two-fold symmetry when rotating the in-plane magnetic field, which is surprising considering the three-fold symmetry of the lattice. This, together with the invariance of the Knight shift upon crossing $T_c$ under an out-of-plane field, indicates the Cooper pairs are in a pseudospin-triplet state with a pinned *d*-vector direction [11]. Subsequently, measurements on Cu-, Sr-, and Nb-doped $Bi_2Se_3$ superconductors all revealed two-fold symmetry in the in-plane field angular dependence of the specific heat, upper critical field, critical current, and magnetoresistance below $T_c$ (Refs. [12-15]). These remarkable observations in $Cu_xBi_2Se_3$ have established it as an odd-parity TSC state [16-18], with the *d*-vector pinned to a specific lattice axis by strong spin-orbit coupling. However, contradictions still exist, especially in the gap symmetry and the pinning directions of the *d*-vectors among the reported bulk and/or macroscopic measurements [11-15]. The *d*-vector is reported to be pinned along the Se-Se bond direction in an NMR study by Matano *et al.* [11], but it lies along the dihedral mirror plane of the Se lattice in specific heat measurements by Yonezawa *et al.* [13] and magnetoresistance measurements by Pan *et al.* [12]. Thus, to reach a comprehensive microscopic understanding of this unique state, and to reconcile the previous contradicting reports, we investigate the superconducting properties of $Cu_xBi_2Se_3$ by STM at ultra-low temperature under magnetic fields with variable direction.

## II. EXPERIMENTAL METHODS

The $Cu_xBi_2Se_3$ (x=0.31) single crystals were prepared via an electrochemical intercalation method as described in Ref. [19]. Magnetic susceptibility measurement shows a superconducting transition temperature (Tc) of about 3 K and shielding fraction of about 17% at 1.8 K (See part 1A of Ref. [20]). STM measurements were conducted in a millikelvin STM system with vector magnetic field, at the base temperature of 20 mK. The effective electron temperature ($T_{eff}$) of the system is calibrated to be 310 mK (see part 2 of Ref. [20]). The $Cu_xBi_2Se_3$ samples were cleaved in vacuum at 77 K and immediately transferred into the STM module. PtIr STM tips were used after being treated on an Au (111) surface. The *dI/dV* spectra are collected using a standard lock-in technique with modulation frequency f = 787 Hz and typical modulation amplitude ΔV = 30-50 μV.

$Cu_xBi_2Se_3$ crystals are thin-bar shaped with the sizes of about 1.7×1.3×0.30 $mm^3$ for sample 1 and 2.5×1.7×0.14 $mm^3$ for sample 2. The demagnetization effect due to the irregular sample shape is very weak as discussed in part 1B of Ref. [20], thus it is neglected in our experiments under magnetic fields. For the measurements under $B_{//}$, the possible residual out-of-plane field component is carefully eliminated via vortex mapping (see part 3 of Ref. [20]). Moreover, the influence of in-plane magnetic vortex and spatially inhomogeneity of superconducting gap under $B_{//}$ were considered and discussed in detail in part 4 of Ref. [20].

## III. RESULTS AND DISCUSSION

The cleaved $Cu_xBi_2Se_3$ surface exhibits three types of regions (Fig. 1), two superconducting (SC-I, SC-II) and one non-superconducting (NSC). The statistics on how often these regions are observed in different samples are summarized in Table I of Ref. [20]. The probability of SC regions being found is unexpectedly low, indicative of the intrinsically inhomogeneous superconductivity of $Cu_xBi_2Se_3$ [21].

Figures 1(a) and 1(b) show the surface morphology of the NSC region. Large, flat terraces are observed, whose heights are integer multiples of 0.95 nm, as shown in Fig. 1(e). This height corresponds to the spacing between adjacent Bi-Se quintuple layers based on previous x-ray diffraction of $Cu_xBi_2Se_3$ (Ref. [4]). The terraces are atomically flat with two kinds of intrinsic defects (Fig. 1(b)). Bright dots are most likely intercalated Cu atoms as seen in previous STM studies [10,22], while trefoil defects, commonly observed on $Bi_2Se_3$ surfaces, arise from the substitution of Se atoms by Bi (Ref. [23]). The inset of Fig. 1(b) shows the undistorted hexagonal Se atomic lattice with the lattice constant of about 0.41 nm. Here we define the ***x***-axis as one of the Se–Se bond directions, and the ***y***-axis as perpendicular to ***x*** within the same Se-plane, as indicated in the inset of Fig. 1(b). As for the electronic structure, a Dirac cone-like feature is suggested by linearly dispersing density of states (DOS)) observed in the d*I*/d*V* spectra shown in Fig. 1(d) [23,24], as well as the quasiparticle interference patterns along a step edge shown in part 5 of Ref. [20]. The extracted Dirac point is located at about 370 meV below the Fermi energy ($E_F$). This is consistent with ARPES measurements on bulk $Cu_xBi_2Se_3$ crystals [5], indicating the existence of topologically non-trivial surface states in NSC regions. A nearly flat DOS is observed around $E_F$ (Fig. 1(c)), precluding the existence of a superconducting gap.

Figures 1(f) and 1(g) display the surface morphologies of one kind of superconducting region (SC-I) observed in sample 1. Large terraces are observed here as well, but their heights are integer multiples of 1.56 nm (Fig. 1(j)), much larger than in the NSC regions. Moreover, as shown in Fig. 1(g), the terraces are rough, with many Cu clusters distributed. It appears that the spacing between $Bi_2Se_3$ quintuple layers in this region is expanded considerably by Cu intercalation, which would lead to a more two-dimensional structure. The SC-I region is adjacent to the NSC region shown in Figs. 1(a) and 1(b). Although the surface lattice cannot be directly seen in SC-I region, the orientation of atomic step edges suggests it has the same lattice orientation with the nearby NSC region (see part 1D of Ref. [20] for details). The *dI/dV* spectrum obtained in SC-I regions (Fig. 1(i)) exhibits a similar Dirac-cone like feature, but the Dirac point is now 650 meV below $E_F$. This significant shift to lower energy compared to the NSC region indicates that the SC-I region is heavily doped with electrons. As shown in Fig. 1(h), a fully developed superconducting gap with pronounced coherence peaks is observed at $E_F$, which is consistent with previous reports [10,19]. We note that the heavy electron doping effect may be critical for the origin of superconductivity here. Such a U-shaped gap is spatially homogeneous in SC-I region, and it is insensitive to Cu clusters and step edges, as shown in Figs. 1(k) and S6 in Ref. [20]. Usually for spin-triplet superconductors, such as $Sr_2RuO_4$, the superconductivity is very sensitive to impurities [25]. However, two theoretical studies suggest a different impurity scattering behavior for the unconventional superconductivity resided in a topological insulator, where the pair decoherence introduced by impurity scattering is strongly suppressed by spin-orbital coupling and spin helical nature in intercalated $Bi_2Se_3$ [26, 27].

Besides SC-I regions, we observe another type of superconducting region (referred to here as SC-II) in sample 2, on islands with steep topographic variations as shown in Figs. 1(l) and 1(m). While the NSC and SC-I regions often exhibit large flat terraces with uniform terrace height, SC-II regions exhibit many irregular terraces of several to tens of nanometers width; the terrace height is non-uniform, even with fractional values observed, whose origin are discussed in part 1E of Ref. [20]. However, the terraces are atomically flat – with the atomic lattice shown in the inset of Fig. 1(m), which is the same as that of the NSC region. No clear Dirac-cone-like feature is observed in this region, while a sharp U-shaped superconducting gap appears at $E_F$, as respectively shown in Figs. 1(o) and 1(n). Why superconductivity occurs in these islands is unclear, and we give some discussions in part 1E of Ref. [20].

The U-shaped spectrum suggests the superconducting gap is fully opened at the Fermi surface, however we find that it still has significant broadening beyond thermal effects, which can be reasonably accounted for by *k*-space anisotropy of the gap size [11-15]. We fitted the superconducting spectra by the Dynes formula [28] with a two-fold anisotropic gap function: $\Delta_k=\Delta_0+\Delta_1|\cos\theta|$ and $T_{eff}$ = 310 mK (see part 6 of Ref. [20]). A small Dynes term $\Gamma$ is used to account for finite quasi-particle lifetime broadening [28]. The fit to the gap in the SC-I region shown in Fig. 1(h) yields $\Delta_0$=0.256 meV, $\Delta_1$=0.206 meV and $\Gamma$=0.004 meV; the fit to a typical superconducting spectrum in the SC-II region (Fig. 1(n)) yields $\Delta_0$=0.477 meV, $\Delta_1$=0.103 meV and $\Gamma$=0.001 meV. Evidently, these regions differ greatly in gap size and gap anisotropy ratio. Significant anisotropy of the superconducting gap is expected for the reported nematic superconducting state of $Cu_xBi_2Se_3$ (Refs. [11-15]).

To further examine the superconducting state, we applied an out-of-plane magnetic field and measured the SC-I regions. Figures 2(a) and S9 in Ref. [20] show zero-bias conductance (ZBC) mappings under various $B_\perp$, where typical vortex lattices are revealed. It is notable that the vortices are significantly elongated along the ***y***-axis and exhibit an elliptical shape. Meanwhile, the vortex lattices are also stretched along this direction. Exponential fits to the profiles along the long and short axes of a vortex under $B_\perp$=0.2 T are shown in Fig. 2(b), which give Ginzburg-Landau coherence lengths ($\xi$) of 30.83 nm and 19.70 nm, respectively, a ratio of 1:0.64. Since the shape of the vortex is related to the superconducting gap structure [29], the elongated vortices observed here are consistent with the highly anisotropic superconducting gap ($\Delta_k$) suggested by the fit in Fig. 1(h), where the ratio between the gap maximum and gap minimum in the SC-I region is estimated to be $(\Delta_0+\Delta_1):\Delta_0$=1:0.55. Moreover, upon increasing $B_\perp$, as the vortex density increases, the vortex size and $\xi$ decrease noticeably (Figs. 2(c) and S9(a-c) in Ref. [20]). Such a significant field dependence of the vortex size may suggest multiband effects [30-33], e.g., the $\xi$ of the surface and bulk states of $Cu_xBi_2Se_3$ could be different. We note that a field-dependent coherence length was also observed in $Bi_2Te_3$ films grown on a $NbSe_2$ substrate [34].

Figures 2(d) and 2(e) show the evolution of *dI/dV* spectra taken along the short and long axes of a vortex under $B_\perp$=0.2 T, respectively. Far from the vortex, the superconducting gap is only slightly affected (blue curves in Figs. 2(d) and 2(e)), and it is gradually suppressed upon approaching the vortex core. Within a few nanometers of one vortex core center, weak ZBCPs are observed with a full-width at half-maximum of about 0.29 meV (red curves in Figs. 2(d) and 2(e)). This ZBCP is observed for the first time in vortices of $Cu_xBi_2Se_3$ (Refs. [10,22]). It could be composed of both conventional Caroli–de Gennes–Matricon bound states and the long-predicted Majorana bound state expected at zero energy in 2D TSC. These are hard to distinguish because the energy separation of the bound states $\delta E= \Delta^2/E_F$ (Ref. [35]) is about 0.4 $\mu$eV here, assuming $\Delta_k$=0.5 meV and $E_F$=650 meV. Nevertheless, our observation of a ZBCP is consistent with the prediction of TSC in $Cu_xBi_2Se_3$. As shown in Figs. 2(f), S9(d), S9(e), and S10 in Ref. [20], ZBCPs have also been observed in some vortex cores under $B_\perp$=0.3 T, but their intensities are slightly weaker than under $B_\perp$=0.2 T. The spectrum near zero bias is distorted for $B_\perp$=0.4 T, and the ZBCP is clearly absent for $B_\perp$=0.6 T, as shown in Figs. 2(f) and S9 in Ref. [20]. This behavior resembles the evolution of the Majorana zero-energy mode observed in 5 QL $Bi_2Te_3$ films grown on $NbSe_2$ (Ref. [36]). The weakening and disappearance of a ZBCP in the vortex core under high field is interpreted as a result of the coupling between states in adjacent vortices [36]. When the field is small, the distance between vortices is large compared to the vortex core size, so the interaction between the vortex bound states can be neglected. Upon increasing field, the inter-vortex distance shortens, enhancing these interactions and eventually destroying the Majorana zero mode [36].

To further illustrate the nematic superconductivity in $Cu_xBi_2Se_3$, we apply an in-plane magnetic field and study the superconducting spectra within the same area of SC-I region as a function of θ, defined as the azimuthal angle of $B_{//}$ with respect to the ***x***-axis of the Se lattice (see inset of Fig. 3(a)). Under various $B_{//}$, no obvious in-plane vortices are observed on the surface, and the superconducting spectra are spatially homogeneous (see part 4 of Ref. [20] for more details). This is possibly because the superconductivity on the surface (likely resides in topological surface state) is more robust to the orbital depairing of an in-plane field, so that the bulk vortex lines are pushed away from the surface, as shown theoretically in Ref. [37]. Consequently, the in-plane field did not generate noticeable spatial inhomogeneity on the superconducting spectrum. However, we observed obvious variation of superconducting spectrum with the direction of $B_{//}$. Figure 3(a) shows typical superconducting spectra when applying a $B_{//}$ of 0.5 T at θ = 90°, 40° and 10°. Clearly, the superconducting gap, as estimated from both the leading edge ($\Delta^*_{LE}$) and coherence peaks ($\Delta^*_{CP}$), varies with θ. We use asterisks (*) here to emphasize that these actually constitute weighted averages of $\Delta_k$ under $B_{//}$. The angular dependence of $\Delta^*_{CP}$ and $\Delta^*_{LE}$ based on an extensive dataset is summarized in Fig. 3(b), and exhibits clear two-fold symmetry, with maxima around θ = 90° and 270° (coinciding with the ***y***-axis) and minima near θ = 0° and 180° (coinciding with the ***x***-axis). Figures 3(c) and 3(d) show similar two-fold symmetry for $B_{//}$=1.0 T. In addition, $\Delta^*_{LE}$ along θ = 85°, 40° and 0° decrease monotonically with increasing $B_{//}$, and no crossover is observed between them, as shown in Figs. 3(e) and S11 in Ref. [20]. Our detailed experiments indicate that the superconducting gap ($\Delta^*_{CP}$ and $\Delta^*_{LE}$) of $Cu_xBi_2Se_3$ exhibits two-fold symmetry throughout the studied $B_{//}$ range from 0.25 to 1.5 T. The suppression of the superconducting gap by $B_{//}$ is much faster along the $\boldsymbol{x}$-axis ($\theta$ = 0°) than nearly along the $\boldsymbol{y}$-axis ($\theta$ = 85°), suggesting a smaller upper critical field $B_{c2}$ along the $\boldsymbol{x}$-axis. Thus, a larger $\xi_y$ is evaluated than $\xi_x$ from the Ginzburg-Landau relations $B_{c2}^{x} = \frac{\Phi_0}{2\pi\xi_c\xi_y}$ and $B_{c2}^{y} = \frac{\Phi_0}{2\pi\xi_c\xi_x}$, which is consistent with our observations that the vortex is elongated along the $\boldsymbol{y}$-axis of the lattice shown in Fig. 2(a). Figure 3(f) displays the oscillation amplitudes of $\Delta^*_{CP}$ and $\Delta^*_{LE}$ under various $B_{//}$. The ratios are different for $\Delta^*_{CP}$ and $\Delta^*_{LE}$, but the trend is the same – both increase gradually as $B_{//}$ increases. For $\Delta^*_{LE}$, the oscillation amplitude increases from 8.7% under $B_{//}$ = 0.25 T to 28.9% under $B_{//}$ = 1.5 T, while for $\Delta^*_{CP}$ it changes from 5.1% to 8.1%. The gradual smearing of the gap edge ($\Delta^*_{LE}$) of the DOS under $B_{//}$ is consistent with the Volovik effect – the quasiparticle energy spectrum is Doppler shifted in the mixed state due to a supercurrent flow running around the vortices, which leads to a broadening of the gap edge even for a nodeless superconductor at finite temperature [38-42].

Our main results are summarized in Fig. 4. The data are plotted in polar coordinates and with respect to the underlying Se lattice. Figure 4(a) shows the profile of an elongated vortex core and the two-fold symmetry of $\Delta^*_{LE}$ measured under $B_{//}$=0.5 T for the SC-I region. The long axis of $\Delta^*_{LE}$ is shown by the red dashed line, which actually nearly coincides with the long axis of the vortex, as well as the dihedral mirror plane (dash-dot lines) of the Se lattice. Figure 4(b) shows the angular distribution of $\Delta^*_{LE}$ under $B_{//}$=1.0 T – its long axis is rotated by about 17.5° off the dihedral mirror plane. This can also be readily seen by comparing the data in Figs. 3(b) and 3(d) directly, as in Fig. S14(a) in Ref. [20].

One can retrieve information on the angular distribution of $\Delta_k$ based on the angular dependence of $\Delta^*_{LE}$. Since the orbital depairing effect of $B_{//}$ (in-plane vortex) is not obvious here, the influence of $B_{//}$ on superconductivity mainly includes two effects: the Volovik [38-42] and Zeeman effects. The former has been extensively studied in nodal superconductors theoretically and experimentally [38-40], and leads to a DOS under $B_{//}$ that exhibits characteristic oscillatory behavior with field orientation. At low temperature and low field, as illustrated in Fig. 4(e), more quasiparticles are excited when the field is perpendicular to the gap-minimal direction of $\Delta_k$ than along other directions, because the field-induced supercurrents (which run perpendicular to the field) can lead to the strongest Doppler shift of the quasiparticle energy around the gap minimal region. Thus, the short axis of $\Delta^*_{LE}$ is actually perpendicular to the gap minimal direction, since this field orientation produces the most quasiparticle excitations. Meanwhile, the application of $B_{//}$ will induce Zeeman effects, whose energy scale is about 0.1 meV for $B_{//}$ = 1.0 T, comparable to the minimal gap magnitude (0.256 meV) in the SC-I region. For $E_u$-symmetry pairing, the *d*-vector is expected to be parallel to the maximal gap direction of $\Delta_k$, and it is normal to the plane in which the spins of the equal-spin-paired electrons are aligned [16]. The Zeeman depairing effect is the strongest when $B_{//}$ is parallel to the *d*-vector, because the electron spins may deviate from the original plane under sufficient $B_{//}$ [25]. Therefore, both the Volovik effect and Zeeman effects will suppress $\Delta^*_{LE}$ much more strongly when $B_{//}$ lies along the maximal gap direction of $\Delta_k$. Consequently, the distribution of $\Delta_k$ for the SC-I region can be qualitatively determined – as illustrated by the thick blue curve in Fig. 4(a), the gap maxima are along a Se-Se bond direction, which is rotated 90° from the long axis of $\Delta^*_{LE}$. The tunneling spectra here and the specific heat measured by Yonezawa et al. [13] reflect the *integrated* DOS over the Fermi surface under the influence of $B_{//}$, and they thus exhibit weaker anisotropy than $\Delta_k$, whose anisotropy can be more precisely estimated by fitting the tunneling spectra or via the vortex anisotropy. The blue curve thus gives the first semi-quantitative sketch of an anisotropic nodeless gap structure in $Cu_xBi_2Se_3$. This angular distribution of the $\Delta_k$ is consistent with the orientation of the observed elongated vortices as, empirically, a vortex will extend out in the direction of a gap minimum (assuming only weak Fermi velocity anisotropy) [29].

For the SC-II region, qualitatively similar data were measured on two domains (see Figs. S12 and S13 in Ref. [20]). $\Delta^*_{LE}$ and $\Delta^*_{CP}$ in this region also vary noticeably with field orientation θ, exhibiting two-fold symmetry, although the $\Delta_k$ here are larger than those of the SC-I regions. Figure 4(c) gives the angular dependence of $\Delta^*_{LE}$ in domain A of the SC-II region under $B_{//}$=0.5 T – the long axis of $\Delta^*_{LE}$ lies along the dihedral mirror plane of the Se lattice. According to above analysis, the angular distribution of $\Delta_k$ in domain A is illustrated by the blue curve, with its maxima again along a Se-Se bond direction. Similar to the case of SC-I region, in domain B of the SC-II region, which has the same lattice orientation as domain A, the long axis of $\Delta^*_{LE}$ under $B_{//}$=1.0 T is rotated to an intermediate angle, about 20.7° off another dihedral mirror plane, as shown in Figs. 4(d) and S14(b) in Ref. [20]. The origin of such a rotation under $B_{//}$=1.0 T is unclear and merits future investigation. Our data show that gap nematicity is ubiquitous in all the superconducting domains of $Cu_xBi_2Se_3$, the long $C_2$ axis of $\Delta_k$ is pinned along one of the three equivalent Se-Se bond directions, which may vary for different SC domains (see more discussion in part 9 of Ref. [20]).

In the theory of Fu *et al.* describing the nematic superconductivity in $Cu_xBi_2Se_3$ (Ref. [16]), two types of odd-parity pairing symmetries were proposed, $\Delta_{4y}$ and $\Delta_{4x}$. $\Delta_{4y}$ is nodeless with its gap maxima along a dihedral mirror plane of the Se lattice (***y***-axis), while $\Delta_{4x}$ has its gap maxima along a Se-Se bond direction (***x***-axis) with a pair of point nodes in the *yz*-plane that are protected by mirror symmetry. The $\Delta_{4y}$ gap distribution is supported by specific heat [13] and magnetoresistance measurements [12]. However, our data suggest that the gap maxima of $\Delta_k$ are pinned along one of the three equivalent Se-Se bond directions, consistent with the NMR study [11], and supporting a $\Delta_{4x}$ pairing symmetry without nodes. The differences among bulk measurements reminds us that the summed responses from all domains with different gap amplitudes and orientations may give an average gap maximum rotated from the actual ones, as the multi-domain effect is directly observed in our study on SC-II region (part 9 of Ref. [20]) and previously discussed by Yonezawa *et al.* in Fig. S8 of their paper [13]. Thus, our findings at the atomic scale facilitate an understanding of the complex phenomena reported in bulk measurements on doped $Bi_2Se_3$ materials [11-15], and provide more-direct constraints on theory.

## IV. CONCLUSION

In summary, our data have revealed a unique nematic superconducting state in $Cu_xBi_2Se_3$, and give a semi-quantitative description of its anisotropic gap distribution, which is likely a nodeless $\Delta_{4x}$ pairing symmetry. On the one hand, we have observed a Dirac-cone-like surface DOS in the SC-I regions, a ZBCP in vortex cores under 0.2~0.3 T out-of-plane fields, and a nematic superconducting order parameter with an anisotropic gap and a preferential symmetry axis. A nematic superconducting state with an anisotropic gap has only been predicted for odd-parity superconductivity [16-18], and the preferential direction of the $C_2$ symmetry axis of the gap and its highly nontrivial response to field strength indeed suggest a pseudo-spin triplet pairing [11]. The ZBCP might be the long-sought Majorana bound state, so our findings further support topologically non-trivial superconductivity in this material. On the other hand, many aspects of our data are not readily explained by current theories for TSC. For example, we neither observed in-gap states (Figs. 1(k) and S15 in Ref. [20]) at the step edge as would be expected for a two-dimensional TSC [8,16,43], nor did we observe in-gap states corresponding to Majorana Fermions on the surface as expected for a three-dimensional TSC [16,18]. It is still mysterious why the predicted point nodes appear to be gapped out, and why the long axis of $\Delta^*_{LE}$ rotates under high $B_{//}$. In short, our findings reveal the microscopic behavior of the nematic superconductivity in $Cu_xBi_2Se_3$, which will facilitate an understanding of its extraordinary properties and microscopic pairing mechanism, and more importantly, our data impose strong constraints for further improving theories of topological superconductivity.

**Acknowledgements**

We thank Guoqing Zheng, Jiangping Hu, Fuchun Zhang, Liang Fu, Dunghai Lee, Jing Wang and Darren Peets for helpful discussions. This work is supported by the National Natural Science Foundation of China, National Key R&D Program of the MOST of China (Grants No. 2016YFA0300200, 2017YFA0303004, and 2017YFA0303104), National Basic Research Program of China (973 Program) under grant No. 2015CB921700, Science Challenge Project (grant No. TZ2016004), and Shanghai Education Development Foundation and Shanghai Municipal Education Commission (Chenguang Program). The work at Cologne has received funding from the European Research Council (ERC) under the European Union's Horizon 2020 research and innovation program (grant agreement No 741121) and was also supported by DFG (CRC1238 "Control and Dynamics of Quantum Materials", Projects A04).

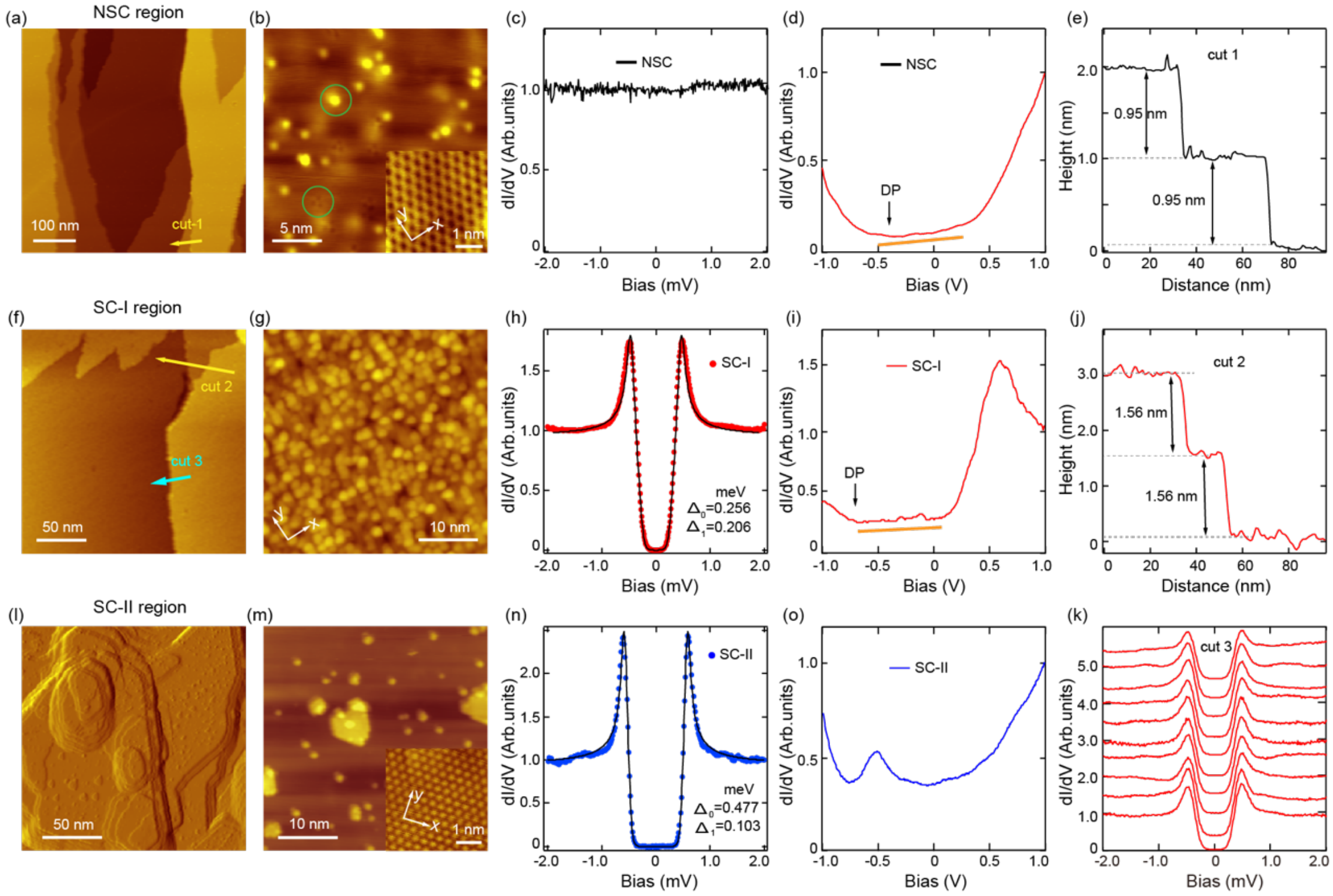

FIG. 1. The three types of $Cu_xBi_2Se_3$ surfaces and corresponding *dI/dV* spectra. (a,b) Typical topographic images of the NSC region ((a): $V_b$ = 3 V, I = 20 pA; (b): $V_b$ = 1 V, I = 20 pA). Two kinds of typical defects are marked by green circles. The inset of (b) shows an atomically resolved image ($V_b$ = 30 mV, I = 70 pA). (c,d) Typical *dI/dV* spectra of the NSC region ((c): $V_b$ = 2 mV, I = 150 pA, $\Delta V$=30 μV; (d): $V_b$ = 1 V, I = 100 pA, $\Delta V$=10 mV). (e) Terrace heights in the NSC region, along linecut 1 in (a). (f,g) Typical topographic images of the SC-I region ((f): $V_b$ = 3 V, I = 10 pA; (g): $V_b$ = 3 V, I = 10 pA). (h,i) Typical *dI/dV* spectra of the SC-I region ((h): $V_b$ = 2 mV, I = 100 pA, $\Delta V$=50 μV; (i): $V_b$ = 1 V, I = 100 pA, $\Delta V$=10 mV). The orange lines in (d) and (i) show the linear dispersion of the DOS; the energies of Dirac points (DP) are indicated. (j) Terrace heights in the SC-I region, along linecut 2 in (f). (k) Superconducting spectra collected across a step edge along linecut 3 in (f). (l,m) Typical topographic image of the SC-II region ((l): differential image, $V_b$ = 3V, I = 20 pA; (m): $V_b$ = 2 V, I = 20 pA). The inset of (m) shows an atomically resolved image ($V_b$ = 5 mV, I = 100 pA). (n,o) Typical *dI/dV* spectra of the SC-II region ((n): $V_b$ = 2 mV, I = 200 pA, $\Delta V$=30 μV; (o): $V_b$ = 1 V, I = 150 pA, $\Delta V$=10 mV).

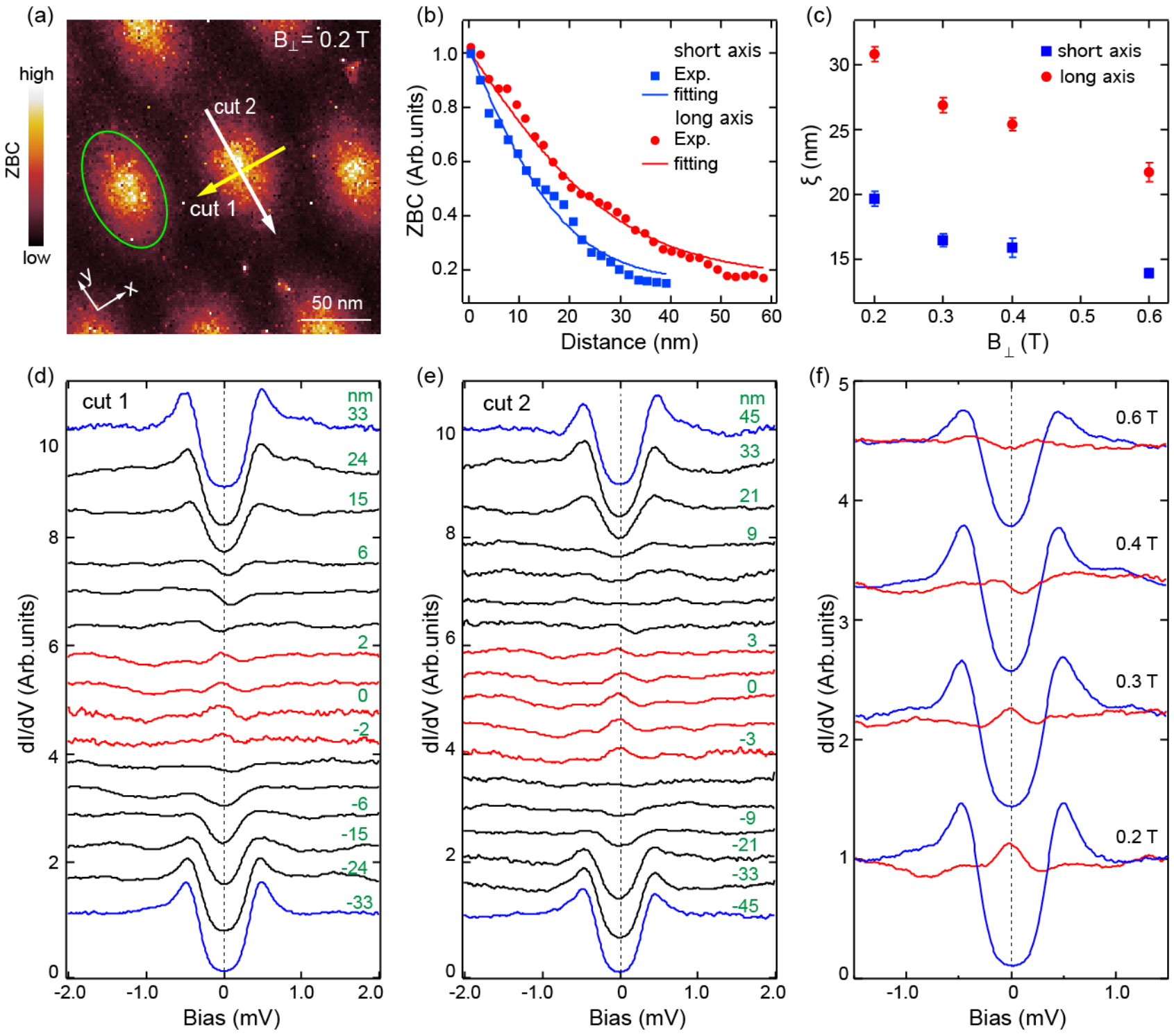

FIG. 2. Vortex state in SC-I region of $Cu_xBi_2Se_3$. (a) Vortex mapping in the SC-I region (225 × 225 nm$^2$) at $V_b = 0$ mV under $B_\perp = 0.2$ T. The profile of one vortex core is indicated by the green ellipse. (b), Exponential fits to line profiles of a single vortex along its long and short axes. (c) Field dependence of the coherence lengths along the long and short axes of the vortex. (d,e) Evolution of the *dI/dV* spectra ($V_b = 2$ mV, I = 100 pA, $\Delta V$=50 μV), taken along the two orthogonal linecuts in (a) as indicated. The green numbers indicate the distances of the spectra from the vortex core center. (f) *dI/dV* spectra as a function of field strength, comparing the vortex core center (red curve) with the bulk tens of nanometers away.

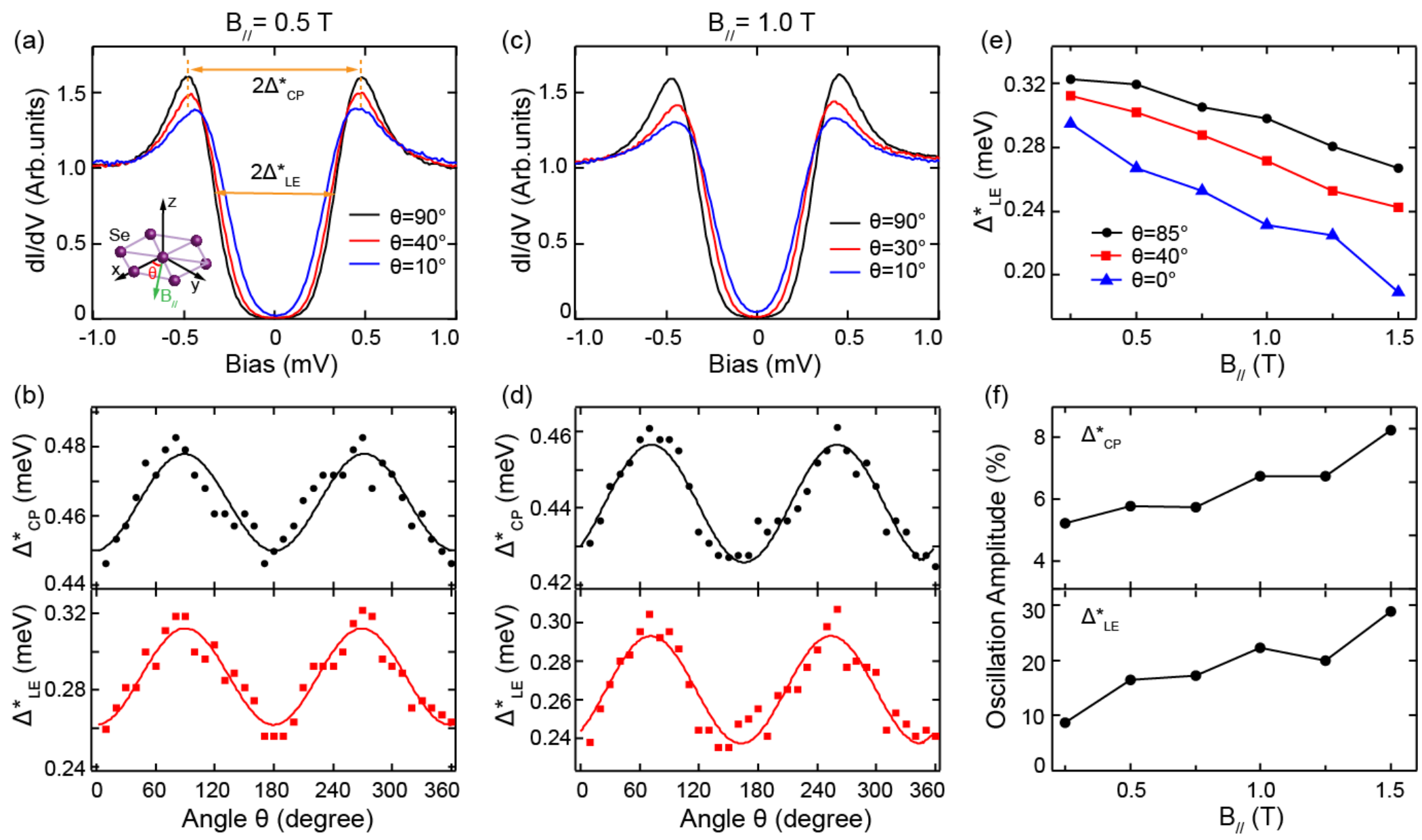

FIG. 3. The average superconducting gap as a function of θ in the SC-I region. (a) Typical *dI/dV* spectra measured under $B_{//}$=0.5 T with different field orientations. The definition of θ is indicated in the inset, where the purple spheres represent Se atoms. The definitions of $\Delta^*_{CP}$ and $\Delta^*_{LE}$ are indicated by the double-arrowed lines. (b) Angular dependence of $\Delta^*_{CP}$ and $\Delta^*_{LE}$ under $B_{//}$=0.5 T. The sinusoidal fit highlights the clear two-fold symmetry. (c,d) Same as (a) and (b) except that $B_{//}$ is increased to 1.0 T. (e) In-plane magnetic field dependence of $\Delta^*_{LE}$ for θ = 85°, 40° and 0°. (f) Field dependence of the oscillation amplitudes of $\Delta^*_{CP}$ and $\Delta^*_{LE}$. The oscillation amplitude is defined here as the ratio of the peak-to-peak amplitude of the oscillation to the maximal gap size at that field.

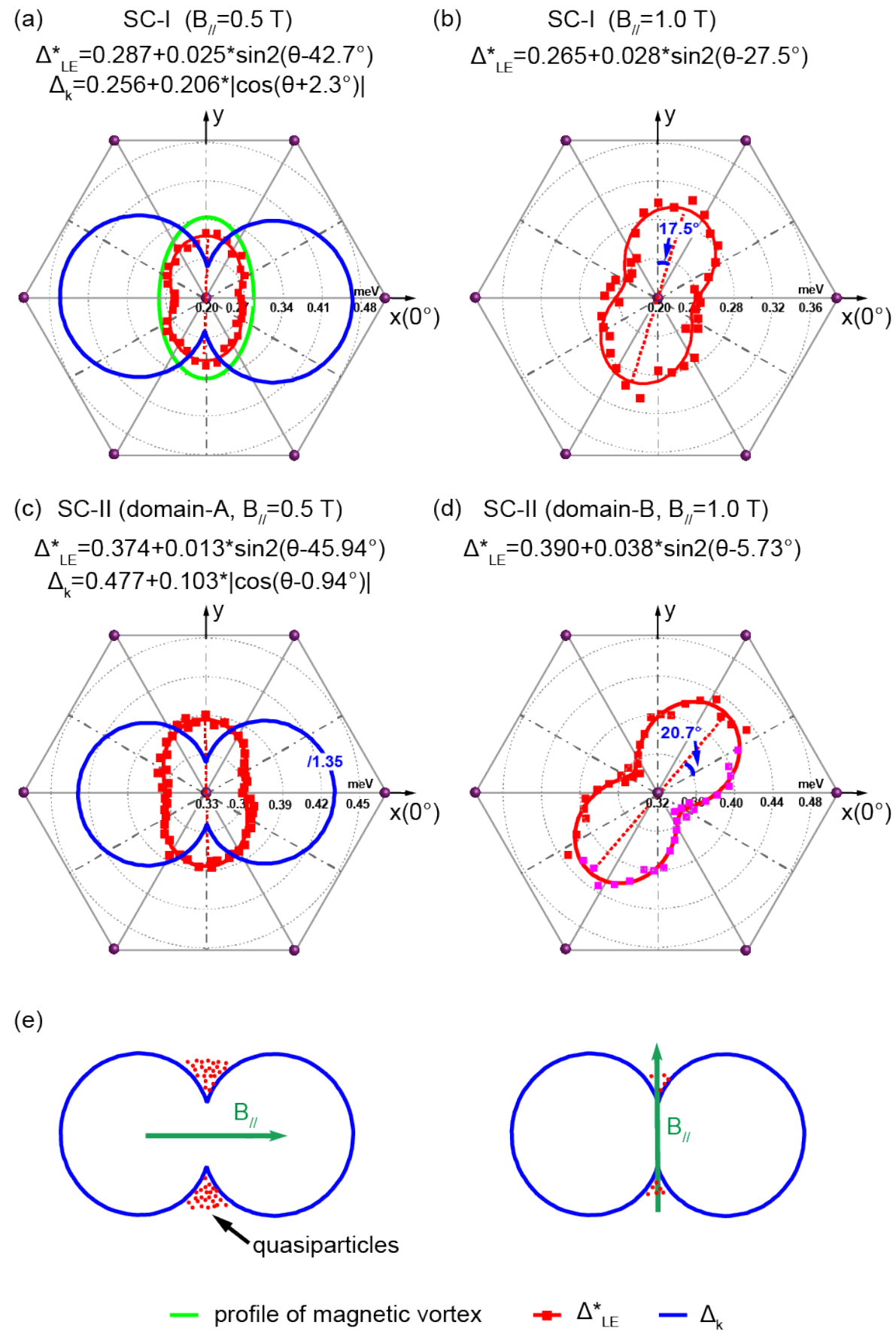

FIG. 4. Nematicity of the superconducting gap in different domains. (a) Comparison of the profile of elliptical vortex and two-fold symmetry of $\Delta^*_{LE}$ under $B_{//}$=0.5 T in the SC-I region, with respect to the Se lattice (the purple spheres represent Se atoms, gray solid lines and dash-dot lines indicate the Se-Se bond directions and the dihedral mirror planes, respectively.). (b) Two-fold symmetry of $\Delta^*_{LE}$ under $B_{//}$=1.0 T in the SC-I region. (c,d) Two-fold symmetry of $\Delta^*_{LE}$ in the SC-II region, observed in domain A under $B_{//}$=0.5 T and domain B under $B_{//}$=1.0 T, respectively. Red squares are the experimental data, while the magenta symbols in (d) denote the symmetric data obtained by shifting the experimental data by 180°. Red dashed lines show the long axis of $\Delta^*_{LE}$ through its two gap maxima. Polar coordinates are plotted out by gray dashed circles, on a scale that differs among the panels; the sinusoidal fit functions describing the two-fold symmetry of $\Delta^*_{LE}$ are also listed to show the oscillation amplitudes and the directions of the $C_2$ axes. The rotation of the long $C_2$ axis of $\Delta^*_{LE}$ under $B_{//}$=1.0 T is marked in (b) and (d). The angular distribution of $\Delta_k$ in various superconducting regions is sketched by thick blue curves, whose anisotropies are extracted from the fitted gap anisotropy in Figs. 1(h), 1(n) and S8. The fitting functions for $\Delta_k$ are also listed. (e) Schematics of the in-plane field angular dependent quasiparticle excitation (red dots) in the low-temperature and low-field condition [39].